\documentclass[10pt]{article}
\usepackage{graphicx}

\def\Title#1{\begin{center} {\Large #1 } \end{center}}
\def\Author#1{\begin{center}{ \sc #1} \end{center}}
\def\Address#1{\begin{center}{ \it #1} \end{center}}

\newcommand\pubblock{\rightline{\begin{tabular}{l} Proceedings of the Fifth Annual LHCP\\ \pubnumber\\
         \pubdate  \end{tabular}}}

\newenvironment{Abstract}{\begin{quotation} \begin{center} 
             \large ABSTRACT \end{center}\bigskip 
      \begin{center}\begin{large}}{\end{large}\end{center} \end{quotation}}

\newenvironment{Presented}{\begin{quotation} \begin{center} 
             PRESENTED AT\end{center}\bigskip 
      \begin{center}\begin{large}}{\end{large}\end{center} \end{quotation}}

\def\beq{\begin{equation}}
\def\eeq#1{\label{#1}\end{equation}}
\def\eeqn{\end{equation}}

\def\beqa{\begin{eqnarray}}
\def\eeqa#1{\label{#1}\end{eqnarray}}
\def\eeqan{\end{eqnarray}}

\let\bar=\overbar

\def\Dslash{\not{\hbox{\kern-4pt $D$}}}
\def\dslash{\not{\hbox{\kern-2pt $\del$}}}

\def\msb{{\bar{\ssstyle M \kern -1pt S}}}

\usepackage{color}

\newcommand\pubnumber{ }

\newcommand\pubdate{\today}

\def\affiliation{
On behalf of the CMS Experiment, \\
INFN - Sezione di Bologna \\
40127, Bologna, Italy}

\begin{document}

% large size for the first page
\large
\begin{titlepage}
\pubblock

%% Change the title, name, abstract
%% Title 
\vfill
\Title{  SEARCHES FOR BSM HIGGS BOSONS IN FERMION DECAY CHANNELS WITH CMS  }
\vfill

%  if you need to add the support use this, fill the \support definition above. 
%   \Author{ FIRSTNAME LASTNAME \support }
\Author{ GIANNI MASETTI  }
\Address{\affiliation}
\vfill
\begin{Abstract}

Recent results on searches for Beyond Standard Model production of Higgs bosons in fermion 
decay channels are presented.
The analyses are based on proton-proton collision data recorded by the CMS experiment at 7, 8, and 13 TeV centre-of-mass energies. 
The exclusion limits determined by the null results of the searches are interpreted in the framework 
of models that include extensions of the standard Higgs sector.

\end{Abstract}
\vfill

% DO NOT CHANGE 
\begin{Presented}
The Fifth Annual Conference\\
 on Large Hadron Collider Physics \\
Shanghai Jiao Tong University, Shanghai, China\\ 
May 15-20, 2017
\end{Presented}
\vfill
\end{titlepage}
\def\thefootnote{\fnsymbol{footnote}}
\setcounter{footnote}{0}
%

% normal size for the rest
\normalsize 

\section{Introduction}

On July 2012 the ATLAS and CMS collaborations announced the discovery of a new particle \cite{Chatrchyan:2012ufa} \cite{Aad:2012tfa} ,
that has been identified as a Standard Model (SM) Higgs boson, $H^{0}$.

In this note the results of searches for additional Higgs bosons performed with
data recorded by the CMS detector \cite{Chatrchyan:2008aa} are summarized.

The searches are generally performed in a model independent approach, and exclusion limits 
in terms of production cross section times the corresponding decay branching ratio are presented. 
The results are interpreted according to beyond-SM (BSM) Higgs scenarios, 
which include the Two Higgs Doublet Model (2HDM) and Triplets Models.

The 2HDM models predict the existence of 5 Higgs Bosons: the three neutral \textrm{\textit{h}}, H, and A, and the charged $H^{\pm}$.
The free parameters can be chosen to be $m_{h}$, $m_{A}$, $m_{H}$, $m_{H^{\pm}}$, the mixing angle $\alpha$ of $M^{2}$, 
the ratio of the vacuum expectation values of the two doublets, $\tan\beta$, and the soft $Z_{2}$ breaking mass parameter $m_{12}$. 
Different 2HDM types can be chosen, depending on the way the two $SU(2)_{L}$ doublets are coupled to the fermion sector. 
In 2HDM of type-I, the $SU(2)_{L}$ doublets couple to both up- and down-type fermions equally; 
in 2HDM of type-II one doublet couples exclusively to up-type and the other exclusively to down- type fermions. 

\begin{figure}[htb]
\centering
\includegraphics[width=0.45\textwidth,trim=0 0 0 0,clip]{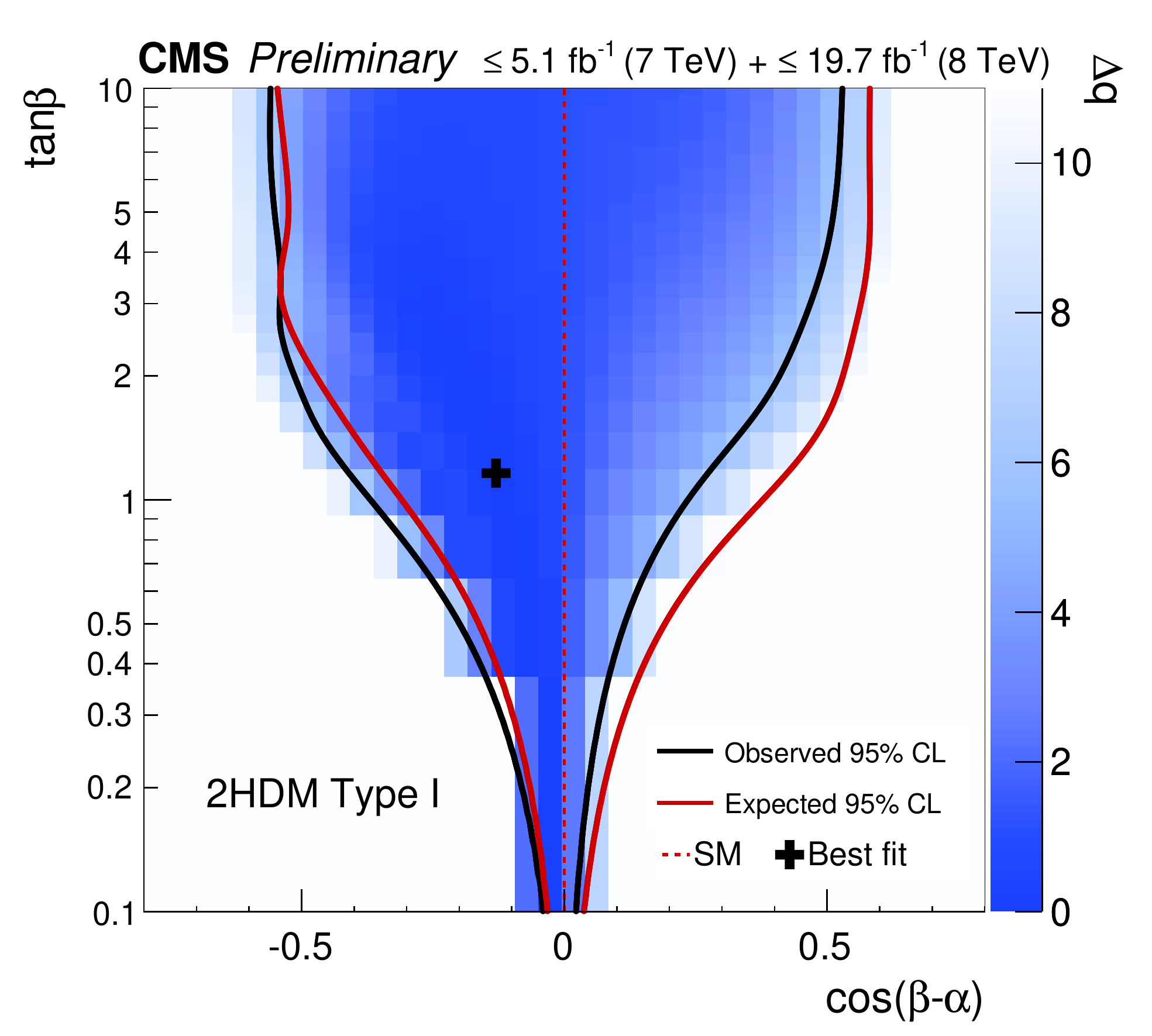}
\qquad
\includegraphics[width=0.45\textwidth,trim=0 0 0 0,clip]{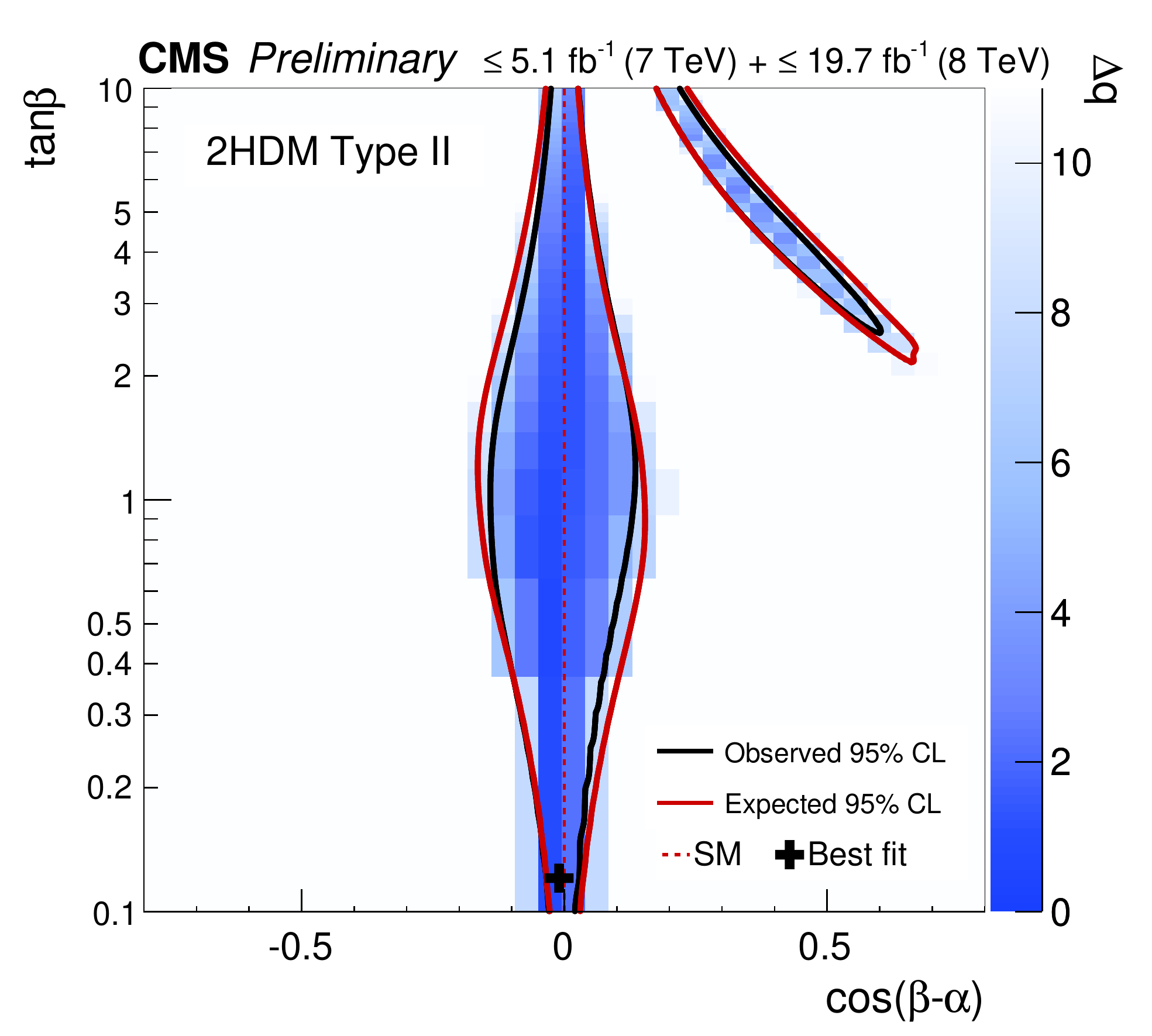}
\caption{General constraints on the 2HDM parameter space obtained from the compatibility with the observed couplings of the $H^{0}$
when interpreted as the \textrm{\textit{h}}. The lines show the contours which restrict the allowed parameter space at the 95\% CL for a 2HDM of (left) Type-I and (right) Type-II. 
 The observed constraints are shown in black. The expected constraints assuming just the SM Higgs sector are indicated by the red continuous line.}
\label{fig:ind_search1}
\end{figure}

Constraints on the angles $\alpha$ and $\beta$ can be set if $H^{0}$ is interpreted as the \textrm{\textit{h}} boson. 
Such constraints have been obtained using the CMS inputs to the combined ATLAS and CMS coupling analysis as presented in Ref.\cite{Higgs_constrains}.
The 95$\%$ CL exclusion contours in the 2HDM of type-I and II, in the $\cos(\beta - \alpha )$-$\tan\beta$ plane, 
are shown in Figure \ref{fig:ind_search1} as obtained from the observed couplings of the discovered $H^{0}$ boson \cite{Higgs_constrains2}.

\section{Neutral Higgs Bosons Searches}

Since LHC Run I, CMS have performed searches for additional neutral Higgs bosons as foreseen in the Minimal Supersymmetric 
Standard Model (MSSM), a particular 2HDM type-II model with two main parameters, $m_{A}$ and $\tan\beta$.
Three possible fermionic final states have been considered: $\tau\tau$ \cite{htau}, $bb$ \cite{hb} and $\mu\mu$ \cite{CMS:2015ooa}.

\begin{figure}[htb]
\centering
\includegraphics[height=3.in]{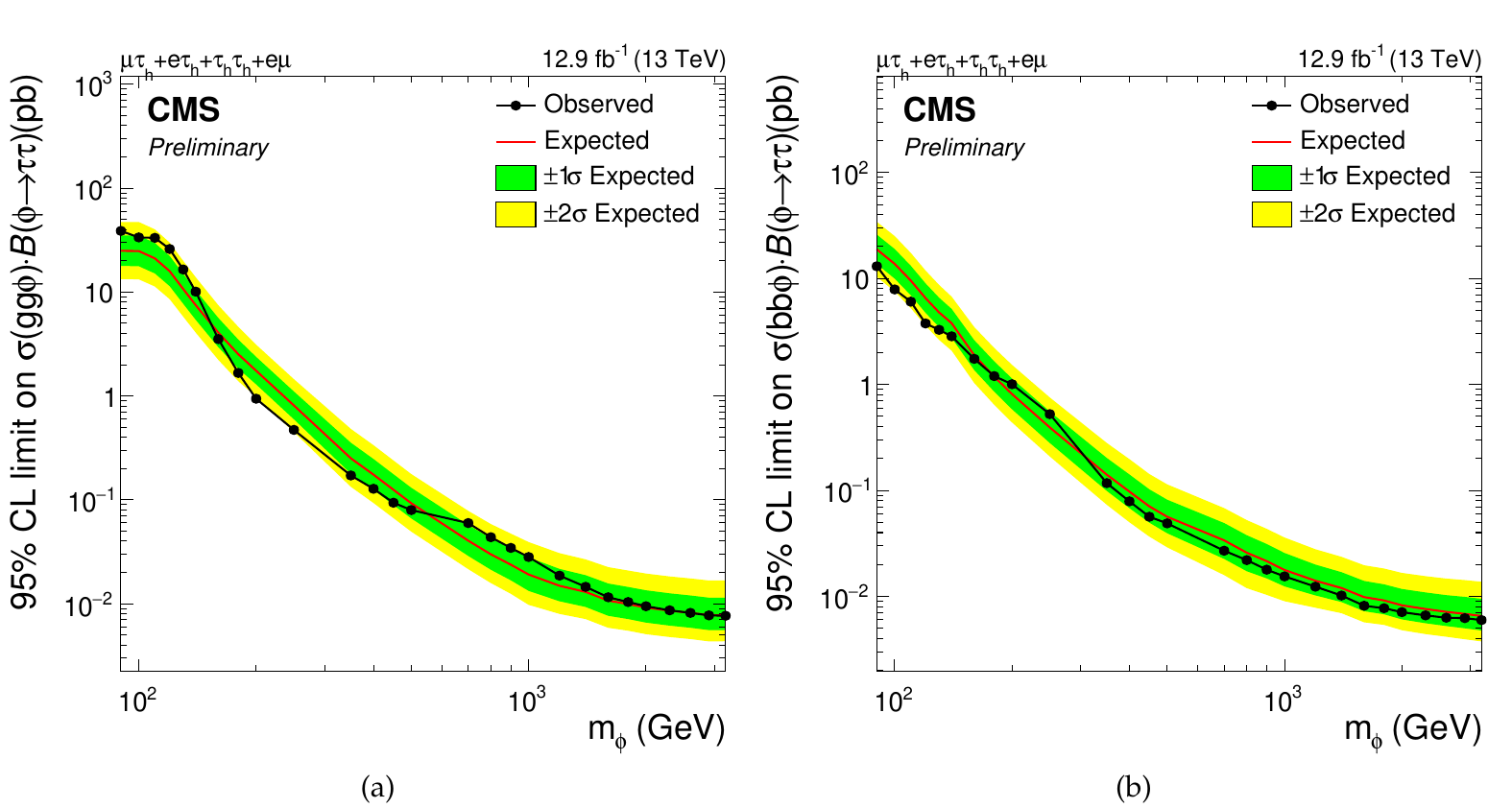}
\caption{Expected and observed limits on cross-section times branching fraction for (a) the gluon fusion process ($gg\phi$) 
and (b) the b-associated production process ($bb\phi$).}
\label{fig:Htau_ind}
\end{figure}

\begin{figure}[htb]
\centering
\includegraphics[width=0.45\textwidth,trim=0 0 0 0,clip]{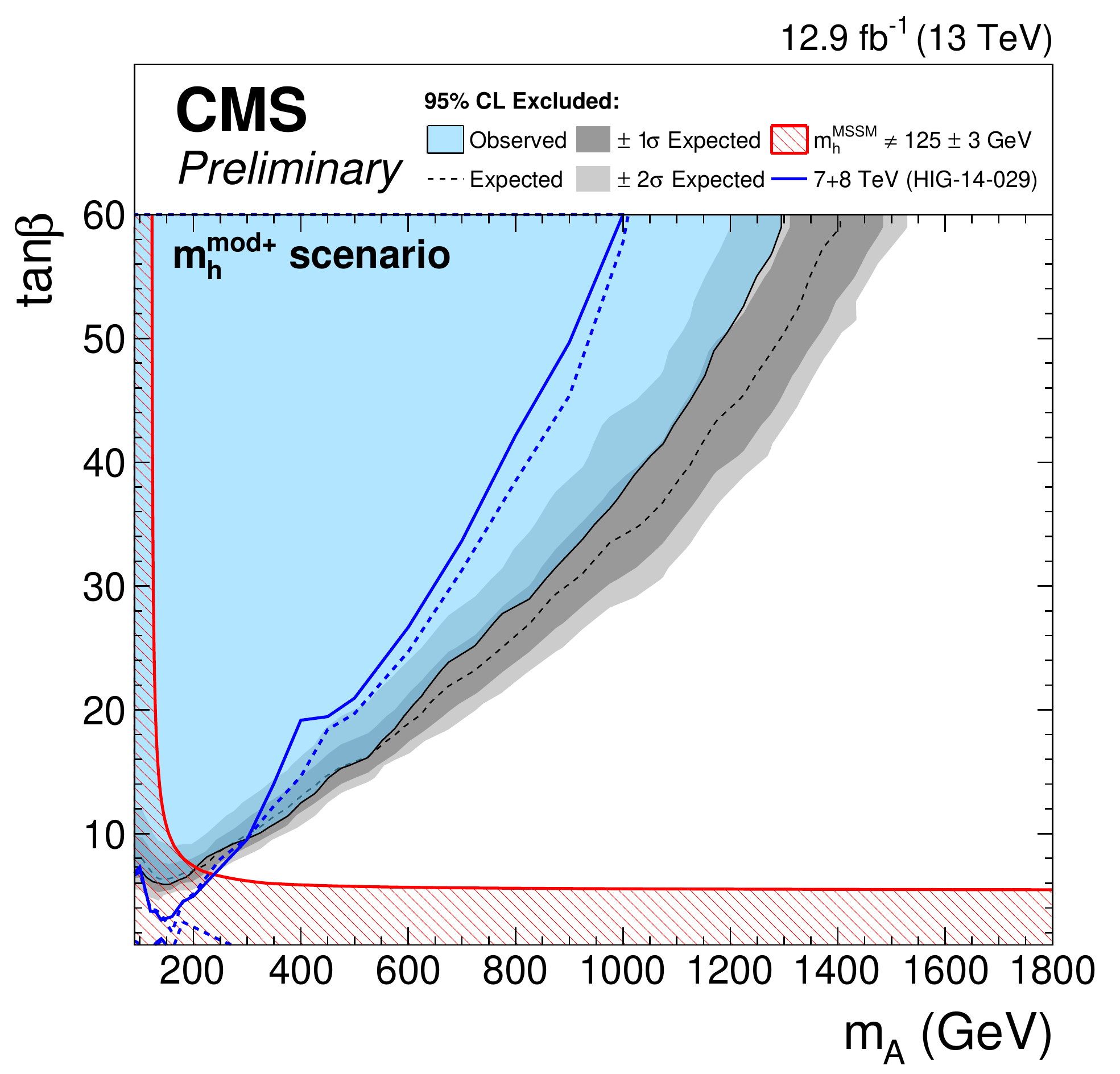}
\qquad
\includegraphics[width=0.45\textwidth,trim=0 0 0 0,clip]{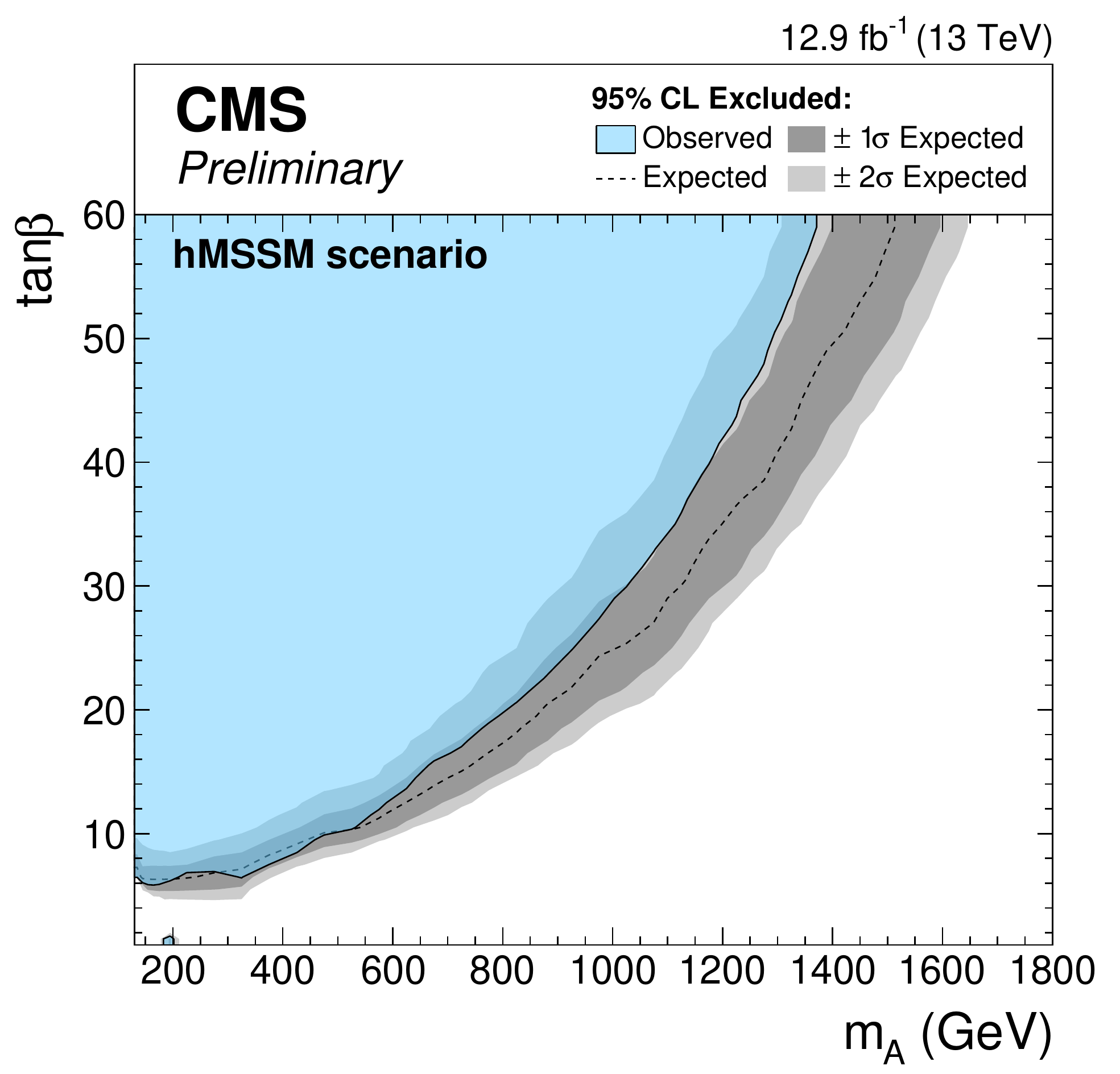}
\caption{Model dependent exclusion limits in the $m_{A}$-tan$\beta$ plane, combining all channels, for (left) the $m_{h}^{mod}$ 
and (right) hMSSM scenarios. 
In the left plot the red contour indicates the region which does not yield a Higgs boson consistent with a mass of 125 GeV 
within the theoretical uncertainties of $\pm$3 GeV, while the blue lines indicate the expected (dashed) and observed (solid) 
exclusions obtained from the most recent Run 1 CMS search for $\phi\rightarrow\tau\tau$ \cite{Khachatryan:2014wca}.}
\label{fig:ind_search3}
\end{figure}

The search of neutral MSSM Higgs via $\tau\tau$ decay has been performed by the CMS experiment considering four decay modes 
of $\tau\tau$ pair: $\tau_{h}\tau_{h}$, $\mu\tau_{h}$, $e\tau_{h}$ and $e\mu$.
No excess above the expectation from the standard model is found and upper limits are set on the production cross sections 
times branching fraction for masses between 90 and 3200 GeV (Fig. \ref{fig:Htau_ind}).
Model dependent limits in the MSSM scenarios are shown in Fig. \ref{fig:ind_search3}.

\section{Charged Higgs Bosons Searches}

\subsection{Single Charged Higgs Bosons}

The 2HDM models predict the existence of two opposite charge Higgs bosons degenerate in mass.
If $m_{H^{\pm}} <  (m_{t} - m_{b})$, then the charged Higgs bosons are mainly produced via top quark decay 
and the dominant decay channel is $H^{\pm}\rightarrow\tau^{\pm}\nu_{\tau}$. 
However, in this region of masses, two additional channels have been searched with CMS Run I data: 
$ \mathrm{c}\overline{\mathrm{s}} $  \cite{Khachatryan:2015uua} and $ \mathrm{c}\overline{\mathrm{b}} $ \cite{CMS:2016qoa}.

\begin{figure}[htb]
\centering
\includegraphics[height=3in]{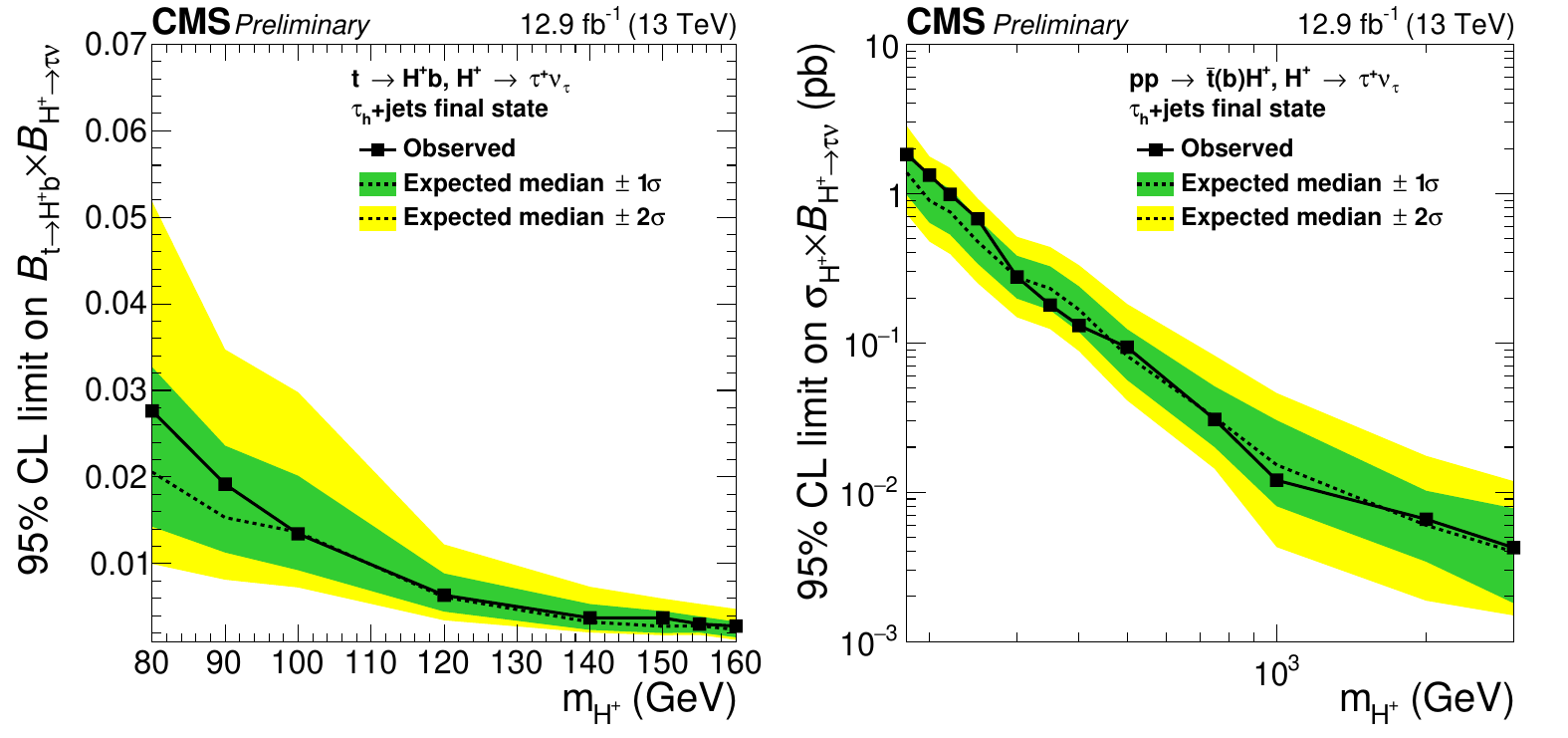}
\caption{The observed 95\% CL exclusion limits (solid points) on 
$\mathcal{B}$(t$\rightarrow$b$H^{\pm}$)$\times$$\mathcal{B}$($H^{\pm}\rightarrow\tau^{\pm}\nu_{\tau}$)
 (left) and $\sigma$(pp$\rightarrow$$H^{\pm}W^{\mp}b\bar{b}$)$\times$$\mathcal{B}$($H^{\pm}\rightarrow\tau^{\pm}\nu_{\tau}$)
 (right) is compared to the expectations from the SM model (dashed line). 
 The green (yellow) error bands represent one (two) standard deviations of the expected limit.}
\label{fig:ChH_ind}
\end{figure}

\begin{figure}[htb]
\centering
\includegraphics[height=3in]{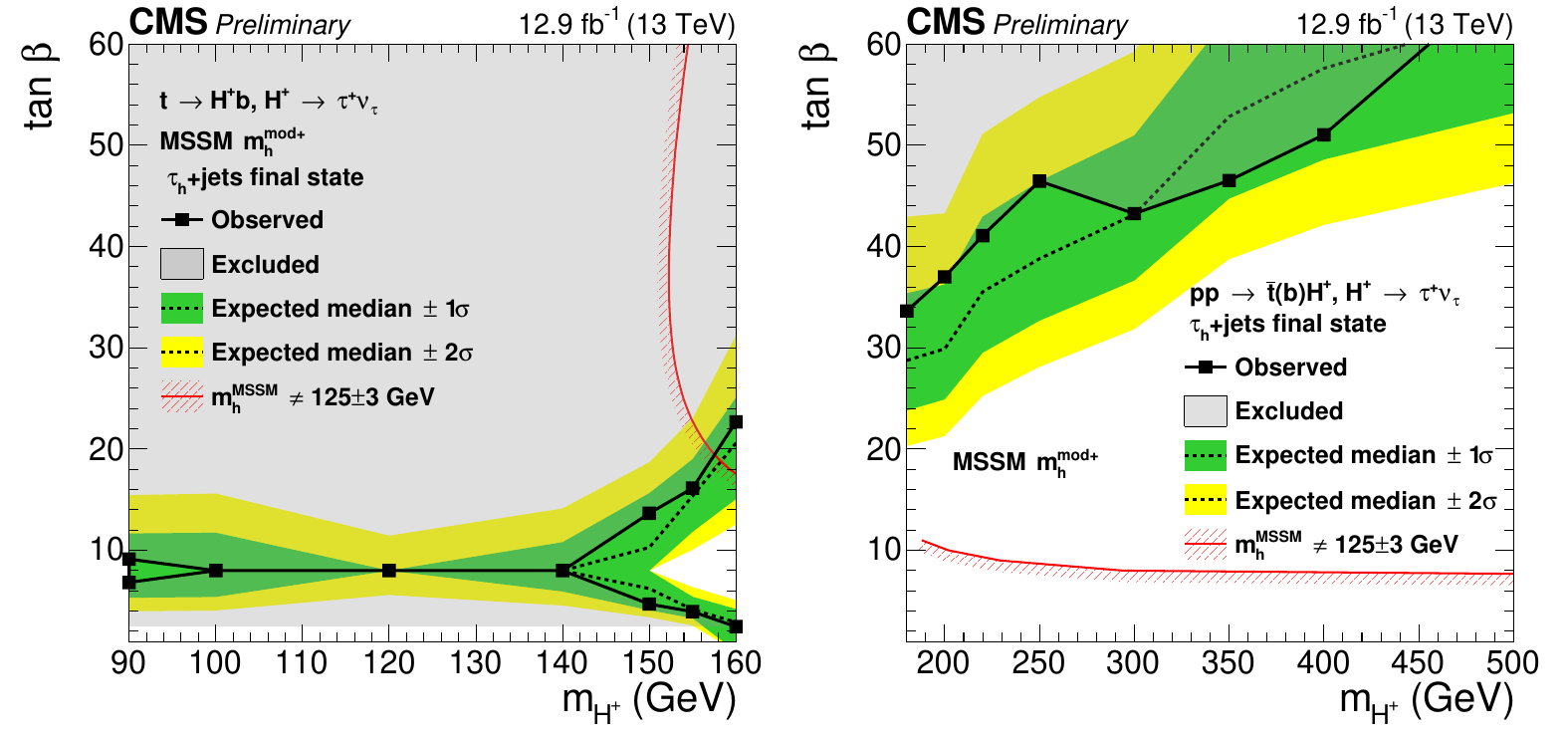}
\caption{Exclusion limits in the $m_{H^{\pm}}$-tan$\beta$ plane in the context of the $m^{mod+}_{h}$ 
model, for the low mass search (left) and the high mass search (right).}
\label{fig:ChH_dir}
\end{figure}

The search in the $\tau^{\pm}\nu_{\tau}$ channel has been recently updated with 12.9~$\mathrm{fb}^{-1}$ of data collected 
at $\sqrt{s} $ = 13~TeV \cite{CMS:2016szv}. The search is performed for masses between 80 and 3000 GeV and 
the fully hadronic final state is considered. 
The observation agrees with the standard model prediction.
In Fig.\ref{fig:ChH_ind} the model independent upper limits are set on the production cross sections 
times branching fraction. 
The results are then interpreted in the MSSM context and the exclusion limits are shown in Fig.\ref{fig:ChH_dir}.

If $m_{H^{\pm}} > (m_{t} - m_{b})$, the $H^{\pm}\rightarrow\mathrm{t}\mathrm{b}$ is expected to be the dominated decay mode.
CMS Run I results can be found in \cite{Khachatryan:2015qxa}.

\subsection{Double Charged Higgs Bosons}

\begin{figure}[htb]
\centering
\includegraphics[height=3in]{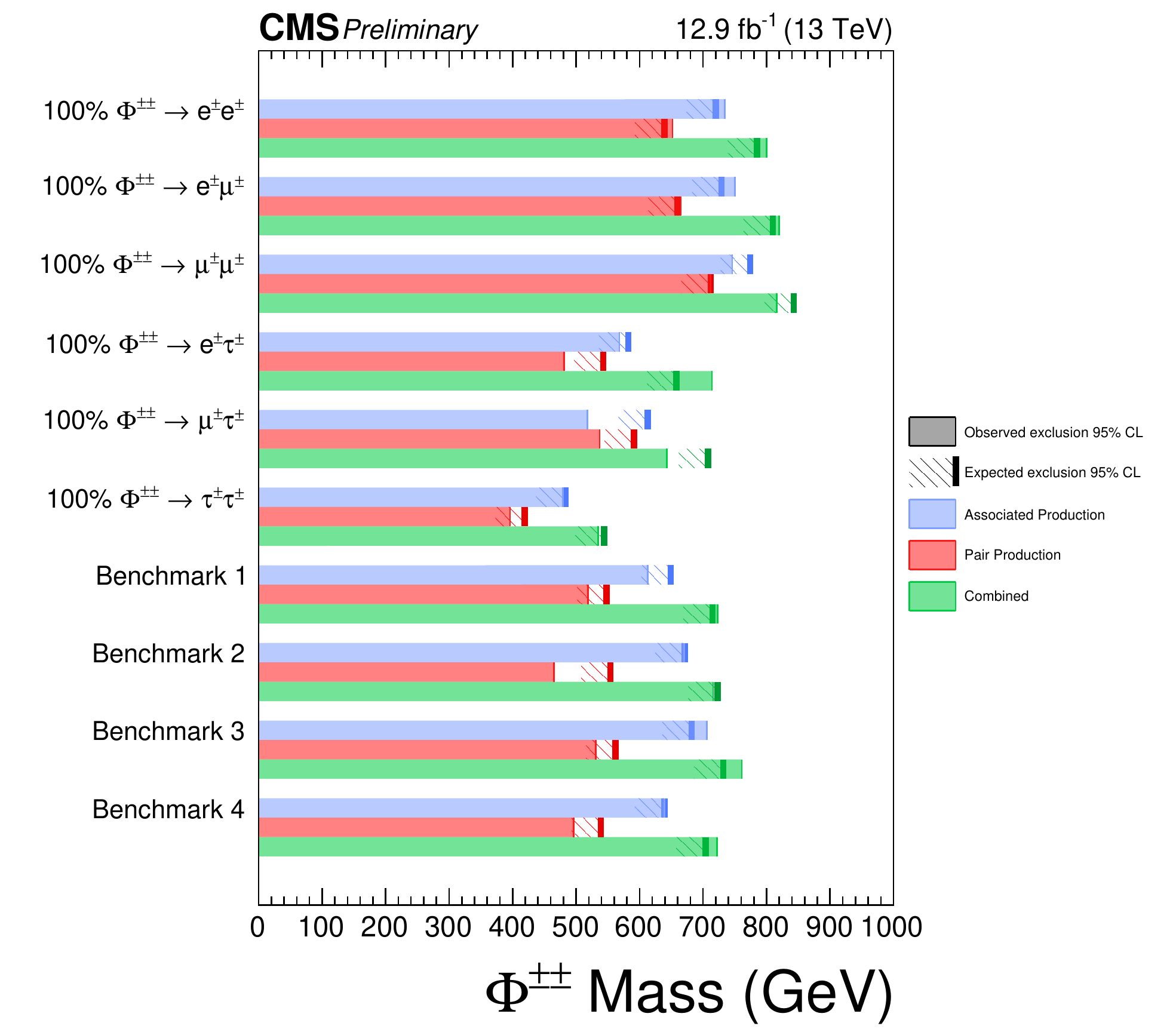}
\caption{Summary of expected and observed limits for each production mode and the combined limit. 
The shaded region represents the excluded mass points and the thick solid line represents the expected exclusion with the hashed region indicating the direction.}
\label{fig:ChH_dir2}
\end{figure}

Double charged Higgs bosons are predicted by models that extend the SM particle spectrum with a scalar triplet.
A search for a doubly-charged Higgs boson, $\Phi^{\pm\pm}$, has been performed with 12.9~$\mathrm{fb}^{-1}$ of data collected 
by the CMS experiment at $\sqrt{s} $ = 13~TeV \cite{CMS:2017pet}. 
The search considers three lepton final states coming from the associated production of 
$\Phi^{\pm\pm}\Phi^{\mp}$ and the four lepton final states coming from the pair production of $\Phi^{++}\Phi^{--}$.
The observation agrees with the standard model prediction. 
Model independent limits are set assuming 100\% decays to single pair of leptons. 
Limits are also set for four benchmarks targeting several neutrino mass hypotheses (Fig.\ref{fig:ChH_dir2}), since 
the Yukawa coupling of the $\Phi^{\pm\pm}$  to leptons is proportional to the mass of the neutrinos. 

\section{Conclusions}

A summary of the BSM Higgs Bosons searches at CMS in fermion decay channels has been presented.
No significant excess to SM prediction has been observed.

All analysis are being updated with full run II dataset.


\begin{thebibliography}{99}

%%
%%  bibliographic items can be constructed using the LaTeX format in SPIRES:
%%    see    http://www.slac.stanford.edu/spires/hep/latex.html
%%  SPIRES will also supply the CITATION line information; please include it.
%%

  
  
%\cite{Chatrchyan:2012ufa}
\bibitem{Chatrchyan:2012ufa} 
  S.~Chatrchyan {\it et al.}  [CMS Collaboration],
  %``Observation of a new boson at a mass of 125 GeV with the CMS experiment at the LHC,''
  Phys.\ Lett.\ B {\bf 716}, 30 (2012)
  [arXiv:1207.7235 [hep-ex]].
  %%CITATION = ARXIV:1207.7235;%%
  %2951 citations counted in INSPIRE as of 22 Jul 2014

\bibitem{Aad:2012tfa} 
  G.~Aad {\it et al.}  [ATLAS Collaboration],
  %``Observation of a new particle in the search for the Standard Model Higgs boson with the ATLAS detector at the LHC,''
  Phys.\ Lett.\ B {\bf 716}, 1 (2012)
  [arXiv:1207.7214 [hep-ex]].
  %%CITATION = ARXIV:1207.7214;%%
  %3009 citations counted in INSPIRE as of 22 Jul 2014
  
  %\cite{Chatrchyan:2008aa}
\bibitem{Chatrchyan:2008aa}
  S.~Chatrchyan {\it et al.} [CMS Collaboration],
  %``The CMS Experiment at the CERN LHC,''
  JINST {\bf 3} (2008) S08004.
  %%  doi:10.1088/1748-0221/3/08/S08004
  %%CITATION = doi:10.1088/1748-0221/3/08/S08004;%%
  %4455 citations counted in INSPIRE as of 22 Aug 2017


 %\cite{Higgs_constrains}
\bibitem{Higgs_constrains}
  CMS Collaboration,
  %``Measurements of the Higgs boson production and decay rates and constraints on its couplings from a combined ATLAS and CMS analysis of the LHC pp collision data at,''
  CMS-PAS-HIG-15-002.
  
  %\cite{Higgs_constrains2}
\bibitem{Higgs_constrains2}
  CMS Collaboration,
  %``Summary results of high mass BSM Higgs searches using CMS,''
  CMS-PAS-HIG-16-007.
  
   %\cite{htau}
\bibitem{htau}
  CMS Collaboration,
  %``Search for a neutral MSSM Higgs boson decaying into $\tau\tau$ with $12.9~\mathrm{fb}^{-1}$ of data at $\sqrt{s}=13~\mathrm{TeV}$},''
  CMS-PAS-HIG-16-037.
  
     %\cite{hb}
\bibitem{hb}
  CMS Collaboration,
  %``Search for a narrow heavy decaying to bottom quark pairs in the 13 TeV data sample,''
  CMS-PAS-HIG-16-025.
    
  %\cite{CMS:2015ooa}
\bibitem{CMS:2015ooa} 
  V.~Khachatryan {\it et al.} [CMS Collaboration],
  %``Search for neutral MSSM Higgs bosons decaying to $\mu^{+} \mu^{-}$ in pp collisions at $ \sqrt{s} =$ 7 and 8 TeV,''
  Phys.\ Lett.\ B {\bf 752}, 221 (2016)
  %%doi:10.1016/j.physletb.2015.11.042
  [arXiv:1508.01437 [hep-ex]].
  %%CITATION = doi:10.1016/j.physletb.2015.11.042;%%
  %16 citations counted in INSPIRE as of 23 Aug 2017
  
%\cite{Khachatryan:2014wca}
\bibitem{Khachatryan:2014wca} 
  V.~Khachatryan {\it et al.} [CMS Collaboration],
  %``Search for neutral MSSM Higgs bosons decaying to a pair of tau leptons in pp collisions,''
  JHEP {\bf 1410}, 160 (2014)
  %doi:10.1007/JHEP10(2014)160
  [arXiv:1408.3316 [hep-ex]].
  %%CITATION = doi:10.1007/JHEP10(2014)160;%%
  %195 citations counted in INSPIRE as of 12 Aug 2017
  
  %\cite{Khachatryan:2015uua}
\bibitem{Khachatryan:2015uua} 
  V.~Khachatryan {\it et al.} [CMS Collaboration],
  %``Search for a light charged Higgs boson decaying to $ \mathrm{c}\overline{\mathrm{s}} $ in pp collisions at $ \sqrt{s}=8 $ TeV,''
  JHEP {\bf 1512}, 178 (2015)
  %%doi:10.1007/JHEP12(2015)178
  [arXiv:1510.04252 [hep-ex]].
  %%CITATION = doi:10.1007/JHEP12(2015)178;%%
  %34 citations counted in INSPIRE as of 24 Aug 2017
  
  %\cite{CMS:2016qoa}
\bibitem{CMS:2016qoa} 
  CMS Collaboration,
  %``Search for Charged Higgs boson to ${\rm c\bar{b}}$ in lepton+jets channel using top quark pair events,''
  CMS-PAS-HIG-16-030.
  %%CITATION = CMS-PAS-HIG-16-030;%%
  %8 citations counted in INSPIRE as of 24 Aug 2017


%\cite{CMS:2016szv}
\bibitem{CMS:2016szv} 
  CMS Collaboration,
  %``Search for charged Higgs bosons with the $\mathrm{H}^{\scriptscriptstyle \pm}\rightarrow \tau^{\scriptscriptstyle \pm}\nu_{\tau}$ decay channel in the fully hadronic final state at $\sqrt{s} = 13~\mathrm{TeV}$,''
  CMS-PAS-HIG-16-031.
  %%CITATION = CMS-PAS-HIG-16-031;%%
  %5 citations counted in INSPIRE as of 24 Aug 2017
  
  %\cite{Khachatryan:2015qxa}
\bibitem{Khachatryan:2015qxa} 
  V.~Khachatryan {\it et al.} [CMS Collaboration],
  %``Search for a charged Higgs boson in pp collisions at $ \sqrt{s}=8 $ TeV,''
  JHEP {\bf 1511}, 018 (2015)
  %%doi:10.1007/JHEP11(2015)018
  [arXiv:1508.07774 [hep-ex]].
  %%CITATION = doi:10.1007/JHEP11(2015)018;%%
  %79 citations counted in INSPIRE as of 24 Aug 2017

%\cite{CMS:2017pet}
\bibitem{CMS:2017pet} 
  CMS Collaboration,
  %``A search for doubly-charged Higgs boson production in three and four lepton final states at $\sqrt{s}=13~\mathrm{TeV}$,''
  CMS-PAS-HIG-16-036.
  %%CITATION = CMS-PAS-HIG-16-036;%%
  %4 citations counted in INSPIRE as of 24 Aug 2017

\end{thebibliography}
\end{document}